\documentclass[useAMS,usenatbib]{mn2e}

\newcommand {\Msol}{M$_{\sun}$}

\newcommand{\fermi}{{\it Fermi }}
\newcommand{\fermilat}{{\it Fermi}-LAT }
\newcommand{\gam}{$\gamma$}

\usepackage{natbib}
\usepackage{longtable}
\usepackage{multicol}
\usepackage{booktabs}
\usepackage{multirow}
\usepackage[tight]{subfigure}
\usepackage{setspace}
\usepackage{epstopdf}
\usepackage[]{graphicx}
\usepackage{booktabs}
\usepackage{amssymb}

\title[A Non-thermal Study of the BCG NGC\,1275]{A Non-thermal Study of the Brightest Cluster Galaxy NGC\,1275 -- The Gamma-Radio Connection Over Four Decades}

\author[K. L. Dutson et al]{K. L. Dutson$^{1}$\thanks{E-mail:
    kate.dutson@leicester.ac.uk}, A. C. Edge$^{2}$,
  J. A. Hinton$^{1}$, M. T. Hogan$^{2}$, M. A. Gurwell$^{3}$ and \newauthor
  W. N. Alston$^{1,4}$\\ $^{1}$Department of
  Physics and Astronomy, The University of Leicester, University Road,
  Leicester, LE1 7RH, UK\\ $^{2}$Department of Physics, Durham
  University, South Road, Durham, DH1 3LE,
  UK\\ $^{3}$Harvard-Smithsonian Center for Astrophysics, 60 Garden
  Street, Cambridge, Massachusetts, MA 02138, USA\\ $^{4}$Institute
  of Astronomy, The University of Cambridge, Madingley Road,
  Cambridge, CB3 0HA, UK}
\begin{document}

\date{Accepted 2014 May 14. Received 2014 May 12; in original form 2014 January 14}

\pagerange{\pageref{firstpage}--\pageref{lastpage}} \pubyear{2014}

\maketitle

\label{firstpage}

\begin{abstract}

Emission from the active nucleus in the core of the brightest cluster
galaxy of the Perseus cluster, NGC\,1275, has varied dramatically over
the past four decades. Prompted by the \fermi detection of flaring in
the \gam-ray band, we present the recent increased activity of this
source in the context of its past radio and \gam-ray output. The broad
correspondence between the high-frequency radio data and the
high-energy (HE) emission is striking. However, on short timescales
this correlation breaks down and the 1.3\,mm Submillimeter Array flux
is apparently unaffected during \fermi-detected flaring activity. The
fact that NGC\,1275 is also detected at TeV energies during the
periods of HE \gam-ray flaring suggests that the short-timescale
variation might be primarily related to changes in the inverse Compton
scattering of photons by the electron population in the jet. The
longer-timescale changes suggest a 30--40 year variation in the
fuelling of the black hole, that affects the power of the inner
jet. NCG\,1275 is a laboratory for the class of brightest cluster
galaxies, and its variability on these timescales has implications for
our understanding of massive galaxies in cooling-core clusters. The
case of NGC\,1275 highlights the need for wide coverage across the
radio band to correctly account for the contribution to emission from
a synchrotron self-absorbed core (for example when considering
contamination of Sunyaev-Zel'dovich effect observations), and the
danger of variability biases in radio surveys of galaxies.

\end{abstract}

\begin{keywords}
galaxies: individual: NGC\,1275 -- galaxies: clusters: individual:
Perseus -- gamma rays: galaxies -- radio continuum: galaxies --
radiation mechanisms: non-thermal -- galaxies: active
\end{keywords}

\section[]{Introduction}
\label{sec:intro}

The impact of an active galactic nucleus (AGN) on its host galaxy and
their wider surroundings was not appreciated until relatively recently
\citep[e.g.][]{bower06,croton06}. The role of AGN feedback in the
formation, evolution, and environment of the most {\it massive}
galaxies is particularly important, given the larger mass of the
central black hole and the denser surrounding gas halo of
these systems, see \citet{mcnamara12} and \citet{fabian12} for recent
reviews.

The prototypical example of AGN feedback within a cluster of galaxies
can be found in the Perseus cluster \citep{bohringer93,fabian01}, host
to the active brightest cluster galaxy (BCG) NGC\,1275. The spectral
properties of the nuclear narrow-line radio source 3C\,84 place it as
intermediate between a BL Lac object and a Seyfert
galaxy. Significantly, NGC\,1275 is one of the very few sources
detected by the {\it Fermi Gamma-ray Space Telescope} (hereafter {\it
  Fermi}), that is not strongly beamed \citep{abdo09a}, and exhibits
pronounced flaring activity \citep[see][]{kataoka10,brown12}.  During
these high-energy (HE) flares, the source becomes significantly harder
\citep{brown12}, and bright enough to detect in TeV \gam~rays
\citep{aleksic12}. The magnitude of the variability in the HE regime
is much greater than that observed in other {\it Fermi}-detected radio
galaxies (e.g. Centaurus-A and M\,87).  Given this, and given the
small sample of such sources, and the fact that NGC\,1275 is the most
massive of these galaxies, particular attention is warranted.

It has been postulated that the central AGN comprises a viscous
accretion disk tilted with respect to the rotating (Kerr)
super-massive black hole, bringing about precessing jets as a result
of torques acting on the disk \citep{falceto10b}. This then accounts
for the observed `S-shaped' morphology of the radio jets, which are
seen to expand into lobes tens of kiloparsecs in extent. Based on
infra-red observations of circumnuclear molecular gas, kinematic
arguments imply a black hole mass of $3.4\times10^{8}$\Msol
\citep{wilman05}. An inherent uncertainty in this value arises from
not knowing the system's true inclination to our line of sight.  Very
Long Baseline Interferometry (VLBI) maps of 3C\,84 can place
constraints on this, suggesting that the milli-arcsecond jet is at
$\leq14.4$\degr~to the observer's line of sight
\citep{krichbaum92}. Closer to the core the constraint is more
stringent ($\theta\leq2.7$\degr), whereas on arcsecond scales the
angle is 39.4\degr$\leq\theta\leq$58.2\degr. For such conditions to be
satisfied, there must be some curvature of the jet away from the line
of sight \citep{dunn06}. \citet{suzuki12} identify new jet components
in the core of NGC\,1275 from Very Long Baseline Array (VLBA) 43\,GHz
imaging, that appear between 2002 and late 2008 and show that the
majority of the increase in flux density at this frequency is due to
the innermost two components. \citet{aleksic13} show that for one of
these VLBA components (`C3'), this value almost doubled between the
end of 2009 and the start of 2011, which can account for the entirety
of the single dish flux increase over this period. Curiously, both
\citet{nagai10} and \citet{suzuki12} claim that C3 is moving relative
to C1, as the latter was in the earliest VLBA observations in
2002. However, an alternative interpretation of these observations is
that C3 is coincident with the nucleus of the galaxy but was
undetected before 2003 due to its inactivity.  We will return to this
point in Section~\ref{ssec:current-outburst}. The core is also known
to vary significantly on short timescales at optical
wavelengths. \citet{kingham79} detected variability of in excess of 30
per cent on timescales of months. Similarly, \citet{lyutyi77} found
variations of 0.2\,mag on 10--20 day timescales. \citet{nesterov95}
show evidence that this optical variability is only observed until
1980.

Observations in the X-ray waveband reveal cavities that anti-correlate
with the radio features, leading to the conclusion that the central
engine is able, through its jets, to inflate bubbles of hot,
relativistic plasma into the surrounding thermal gas of the
intracluster medium (ICM), displacing it. Further evidence to support
this includes the additional X-ray cavities to the north-west and
south of the core, where the the radio lobes have detached and risen
buoyantly through the ICM, forming `ghost bubbles'.  Indeed, the
northern bubble has been likened in appearance to a spherical cap air
bubble rising in water \citep{fabian03b}. The position of the
older generation of X-ray cavities with respect to the younger,
radio-active pair is misaligned, suggesting that the detachment
occurred at a different phase in the cycle of jet precession. The
observed X-ray and radio emission can be produced by a precession
period, $\tau_{{\rm prec}}$ of $3.3 \times 10^{7} {\rm yr}$, an
opening angle of 50\degr, and an inclination of the precession axis to
the line of sight of 120\degr~\citep{dunn06}.

On larger scales, Perseus hosts a $\sim$300\,kpc radio mini-halo
centred on NGC\,1275, with spectral steepening away from the cluster
core. It has been proposed that the energy required for the particle
acceleration is provided by magnetohydrodynamic (MHD) turbulence in
the cooling flow \citep{gitti02}.

Another feature observed in NGC\,1275 is a spectacular optical
nebulosity: a filamentary system of ionised, H$\alpha$-emitting gas
extending over 100\,kpc \citep{fabian03a}. Both radial and tangential
filaments are seen, the latter curving around the galaxy
\citep{conselice01}, and the brighter features are also detected in
the soft X-rays. It has been suggested that they reveal the flow of
the hot gas of the rising radio bubbles. An example of such a tracing
streamline is the `horseshoe'-shaped filament, which clearly turns
back on itself at the boundary of the outer buoyant bubble. One
interpretation then, is that the bubble draws the ionised gas into its
wake as it rises \citep{fabian03b}, and there is enhanced cooling
along its edge, where the ICM is compressed.

The need for a feedback mechanism within the Perseus system to
counteract the cooling flow was identified with such measurements as
the temperature of intracluster gas, which has been found to be above
a third of the virial temperature, given that the ICM has a short
cooling time on the galactic scale \citep{salome06}. The coincidence
of X-ray cavities with the radio lobes of NGC\,1275 is evidence that
the central source can suppress gas cooling. Mechanical feedback from
the AGN, with an energy output estimated to be $\approx
10^{43}$\,erg\,s$^{-1}$, is able to contribute toward this re-heating
effect \citep[see e.g.][]{fabian94}. Whether some additional source of
heating is also necessary to produce the observed suppression is
highly debated. Critically, the AGN output of NGC\,1275 is variable on
decade-long timescales. \citet{odea84} show that NGC\,1275 varied
substantially over the 22 years between 1960 and 1982 at
2--100\,GHz. \citet{abdo09a} note that the 15\,GHz flux density of
NGC\,1275 declined dramatically between 1982 and 2005, but has
increased steadily since then. They also observe that these changes in
radio output are mirrored in the \gam~rays. It is this connection that
we investigate in more detail in this paper.

We begin by describing our analysis procedure in
Section~\ref{sec:analysis}. Our results are presented in
Section~\ref{sec:results} and discussed in
Section~\ref{sec:discussion}, with a focus on the connection between
the non-thermal sub-millimetre emission and the gamma radiation,
including a review of the historical activity of the
source. Conclusions are then drawn and prospects suggested in
Section~\ref{sec:conclusion}.

\section[]{Observations and Data Analysis}
\label{sec:analysis}

On board the \fermi spacecraft is the Large Area Telescope (LAT): a
pair conversion instrument sensitive to high-energy photons
($\sim$\,20 MeV --\,$>$300 GeV) with a large effective area and wide
field of view (2.4 sr, and $\sim$\,6500\,cm$^{2}$ on-axis at 1\,GeV
based on post-launch instrument response functions (IRFs),
respectively). A detailed description of the detector is given in
\citet{atwood09} and references therein, and a post-launch review of
its in-flight performance and the IRFs in
\citet{ackermann12}. Primarily, observations are made when the \fermi
satellite is in {\it sky survey mode}, i.e. the spacecraft is pointed at
some angle from the zenith and rocked about its orbital plane, north
for one orbit then south for the next, providing uniform sky exposure
on a $\sim$3-hour cycle.

Reprocessed {\it sky survey mode} event and spacecraft data amassed
between the 4$^{th}$ of August 2008 (the date after which
science-data-taking began) and the 6$^{th}$ of March 2014 (a
Mission-Elapsed-Time interval of 239557417 to 415756803, equating to
$\sim$2039 days) were extracted for a source region (SR) of
20\degr~centred on the position of NGC\,1275 (RA =49.951\degr,
Dec. =$+41.512$\degr; taken from data in the literature) and in the
energy range $300\,{\rm MeV} \leq E \leq 300\,{\rm GeV}$, which
excludes events at the lowest energies visible to \fermi, where the
high point spread function (PSF) leads to considerable source
confusion. A zenith cut of 100\degr~was applied to exclude HE photons
from the Earth's limb, and all `Source' class events were
accepted. The data were analysed using the \fermi Science Tools
version v9r32p5, and IRF P7REP$\_$SOURCE$\_$V15.

A binned, point-source analysis of NGC\,1275 was carried out. Data
were fitted to a source model constructed to encompass emission from
local (that is, within the SR) sources listed in the \fermilat Second
Source Catalog (2FGL; \citealt{nolan12}), the diffuse Galactic and
extragalactic (isotropic) \gam-ray backgrounds (recent models
gll\_iem\_v05.fits and iso\_source\_v05.txt
respectively\footnote{http://fermi.gsfc.nasa.gov/ssc/data/access/lat/BackgroundModels.html}
were used; their normalisation free to vary) and the candidate source
itself. The photon index and normalisation for objects within
1\degr~of NGC\,1275 were allowed to vary. Outside of this region but
within the specified {\it region of interest} (ROI) of 10\degr, the
normalisation was free to vary, but the photon index was fixed to the
2FGL value. Sources outside of the ROI but within the SR were included
in the model, but both aforementioned parameters were fixed to the
catalogue values.

The maximum-likelihood spectral estimator {\sc gtlike} was used to
perform a {\it binned} fit, modelling the source emission with a power
law spectrum. {\sc gtmodel} was then used to obtain a model map of the
region given the result of the fit. {\sc gtbin} was utilised to
construct a counts map from the data, and a binned exposure cube was
generated (through use of the {\sc gtexpcube2} tool) as part of the
binned likelihood analysis chain. A binned method of likelihood
fitting was adopted over an unbinned one, because it has been found to
be more robust. The fit outputs a Test Statistic for the source, a
spectral index and an integrated flux.

{\sc gtbin} was also used to construct source light curves through
aperture photometry. Monthly, fortnightly, and 3.5-day binning
was applied to the data, within an ROI of 1\degr. The background was
estimated in two manners: firstly, the model-predicted source flux
contained therein was subtracted from the overall model-predicted
flux, calculated by integrating over the ROI in the corresponding
model map given weighted exposure maps across thirty bins in
energy. This value was then multiplied by a correction factor to
account for the tails of the PSF that lie outside the aperture
selected for the light curve, which was estimated by taking the ratio
of the average exposure across all time bins for the light
curve-specific ROI and the larger region used to perform the
likelihood fit, calculated using the {\sc gtexposure} tool. Secondly a
`mean background' light curve was constructed by averaging the flux
within a pair of {\it off-source} regions in the field (at equal
Galactic latitudes to NGC\,1275, and containing no 2FGL sources)
binwise. Subtracting either of these backgrounds from the raw flux in
each bin gives an estimate of the relative excess from the source, and
allows for comparison between model-dependent and independent measures
of the flux level.

The data were divided into two bands in energy (a `low' cut below 3
GeV, and a `high' cut above 3 GeV) and the initial analyses repeated
in an otherwise identical manner. The ROI selected for constructing
light curves in these bands was based on the LAT PSF at the lower
energy bound (0.4\degr\ for the high band; 2.3\degr\ for the low band)
in an effort to conserve all available source photons. The evolution
of hardness ratio (HR) was investigated by dividing the high-band
excess flux by that of the low-band, binwise. The HR was also plotted
against the monthly source flux across the full energy range.

The historical radio data have been collated from a number of
different studies and observatories. The 14.5\,GHz University of
Michigan Radio Astronomy Observatory (UMRAO) monitoring of NGC\,1275
is presented in \citet{abdo09a} and we refer the reader to this paper
and the excellent historical record that the UMRAO project provides
for this source (and many others) for the 4.8, 8.0 and 14.5\,GHz
light curve of NGC\,1275. Our emphasis in this work is the
higher-frequency ($>$20\,GHz) data where the coverage is less
coherent. \citet{odea84} present the best pre-1980 light curve of
NGC\,1275, to which we have added the Mets\"ahovi data
\citep{nesterov95} at 32--37\,GHz and 90\,GHz. We also extract
photometric points from \citet{schorn68}, \citet{hobbs69},
\citet{fogarty71}, \citet{epstein79}, \citet{landau80},
\citet{flett81,flett83}, \citet{barvainis84}, \citet{steppe93},
\citet{reuter97}, and \citet{nagai12}, at both of these frequencies.  We
include yearly averages of the 85--95\,GHz fluxes from
\citet{trippe11} in the period between 1996--2010, which are a subset
of the data presented in that paper.

The flat spectral index of the source at 30--40\,GHz means that the
differences between 32 and 37\,GHz observations are relatively small,
and are swamped by the order-of-magnitude change seen over the last 50
years.

Finally, we include Wilkinson Microwave Anisotropy Probe ({\it WMAP})
and {\it Planck} 33/30 and 94/100\,GHz data, respectively
\citep{wright09,planck11} to show the most recent fluxes averaged over
each of the nine years of the {\it WMAP} mission (August 2001 to
August 2010), and the first 15 months of {\it Planck} operations
(August 2009 to November 2010). 

\begin{figure*}
  \centering
  \includegraphics[width=16cm]{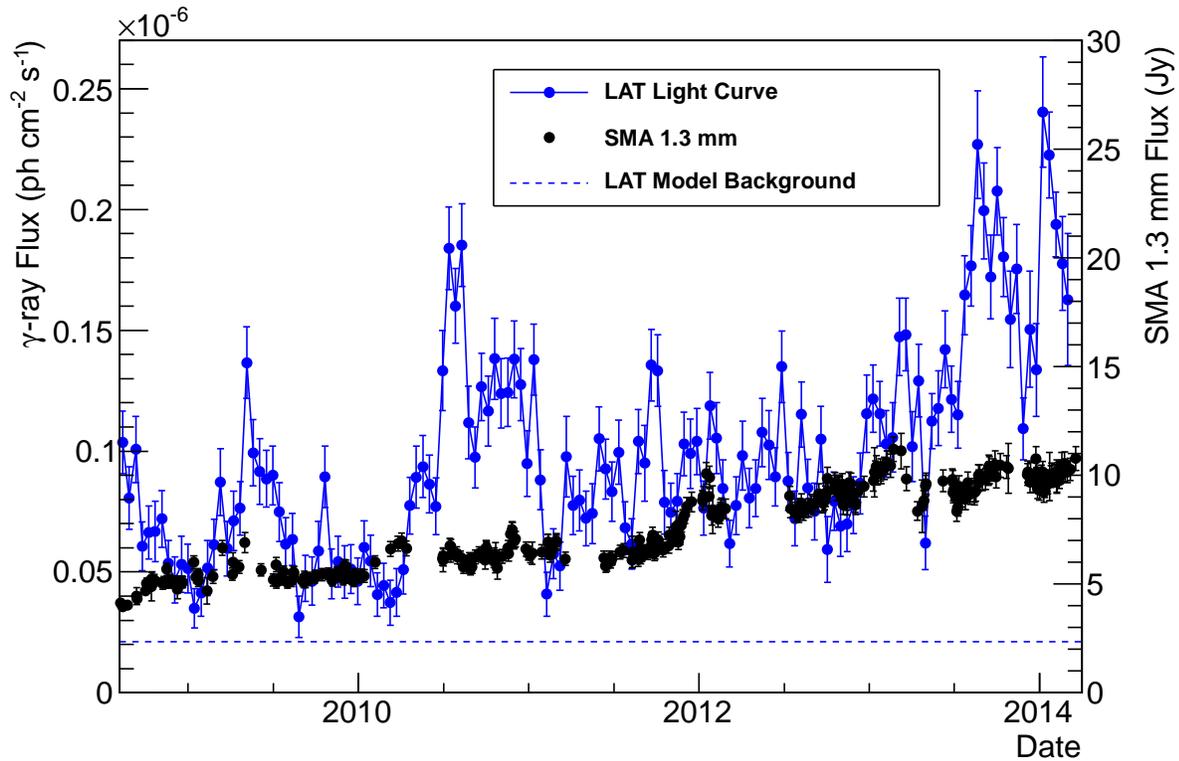}
  \caption{The \fermi light curve for NGC\,1275 since launch (August
    2008), between 300\,MeV and 300\,GeV and within 1\degr, with
    fortnightly binning. The model-predicted background level is
    indicated, and has been subtracted from the raw flux to produce
    the source light curve. Overlaid is the SMA calibrator monitoring
    data at 1.3\,mm. }
  \label{fig:1}
\end{figure*}

We also make use of 1.3\,mm data from the Submillimeter Array (SMA)
calibrator monitoring, to provide a more detailed flux comparison
since the launch of {\it Fermi}, as it has the largest number of
available datapoints. \citet{gurwell07} describe the SMA Phase
Calibrator monitoring in detail. NGC\,1275/3C\,84 has been observed
regularly from the start of SMA operations in late 2002 at 1.3\,mm and
870\,$\mu$m. Save for gaps between March and May each year when the
source is a daytime object, the time coverage is relatively uniform
with typically several observations per week, see Fig.~\ref{fig:1}. We
use all the available data within a frequency range 220--230\,GHz to
ensure the fluxes are comparable. The SMA observations are made at a
variety of baseline configurations, so some flux could be resolved out
at the longest baselines. However, given the radio structure at lower
frequencies, the fraction of the 1.3\,mm continuum that comes from
scales larger than $\sim$2\arcsec~is likely to be small even when the
source is at its weakest. Therefore the SMA flux should be unaffected
by the configuration and any increase can be attributed to flux from
the unresolved core component. This selection makes use of over 80 per
cent of the 1.3\,mm SMA datapoints, as there is a strong preference
for this narrow frequency range from SMA observers.

\section[]{Results}
\label{sec:results}

\begin{figure*}    
  \begin{center}
    \subfigure{%
      \includegraphics[width=15cm]{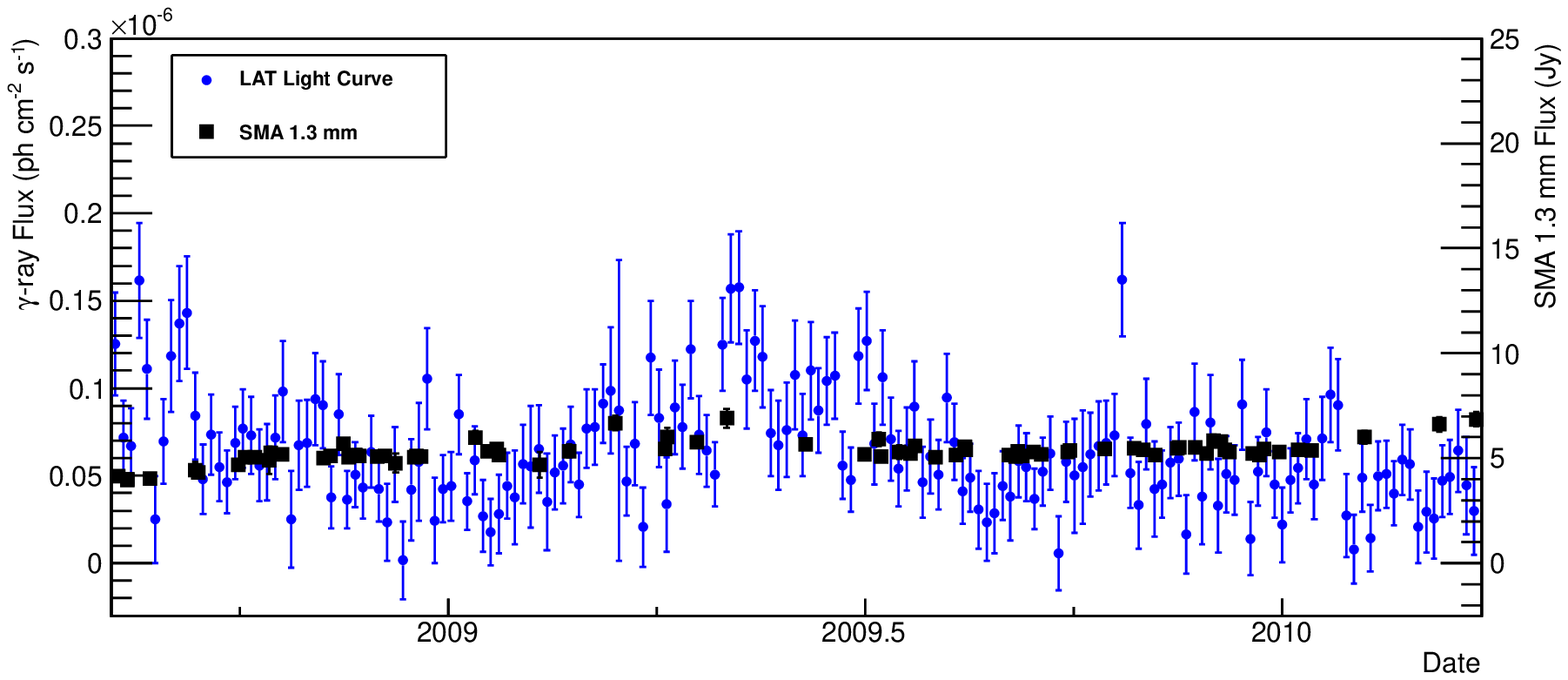}}\\
    \subfigure{%
      \includegraphics[width=15cm]{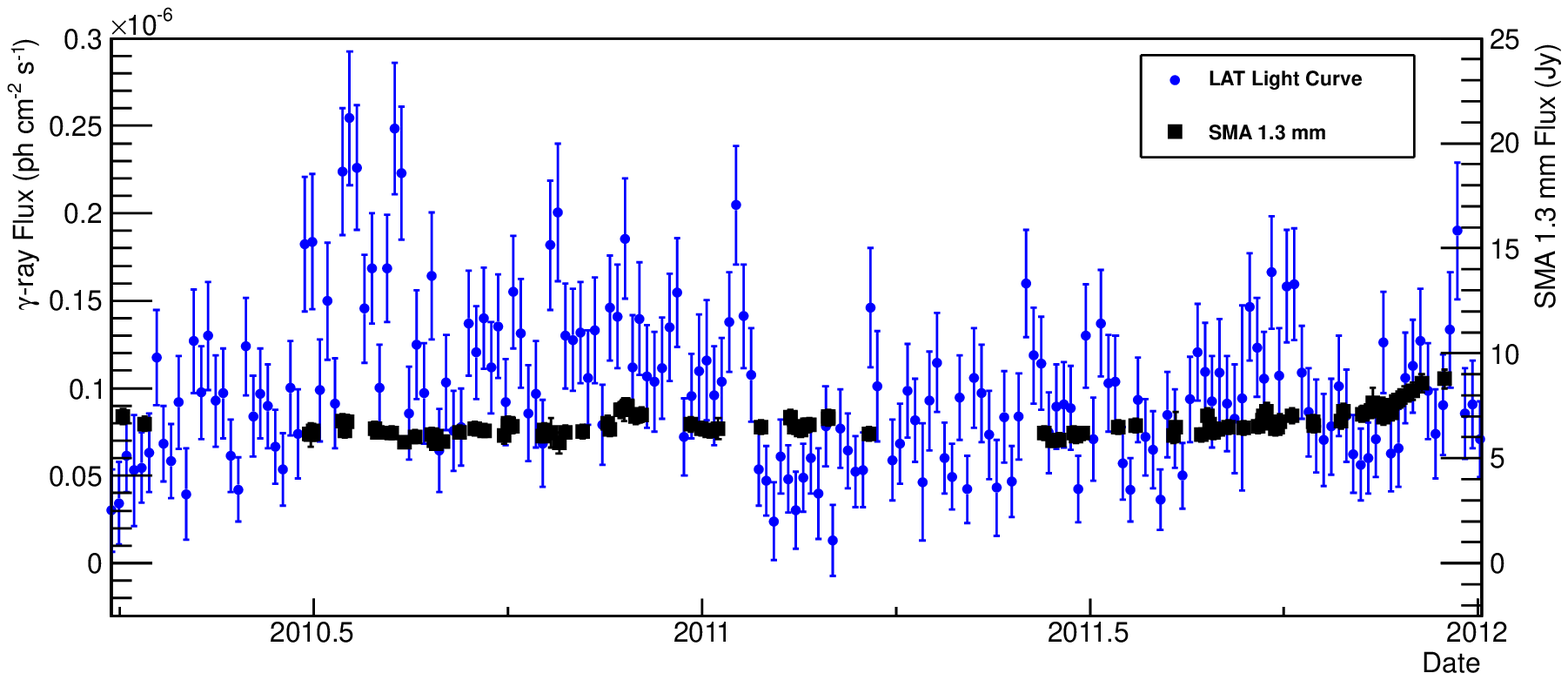}}\\
    \subfigure{%
      \includegraphics[width=15cm]{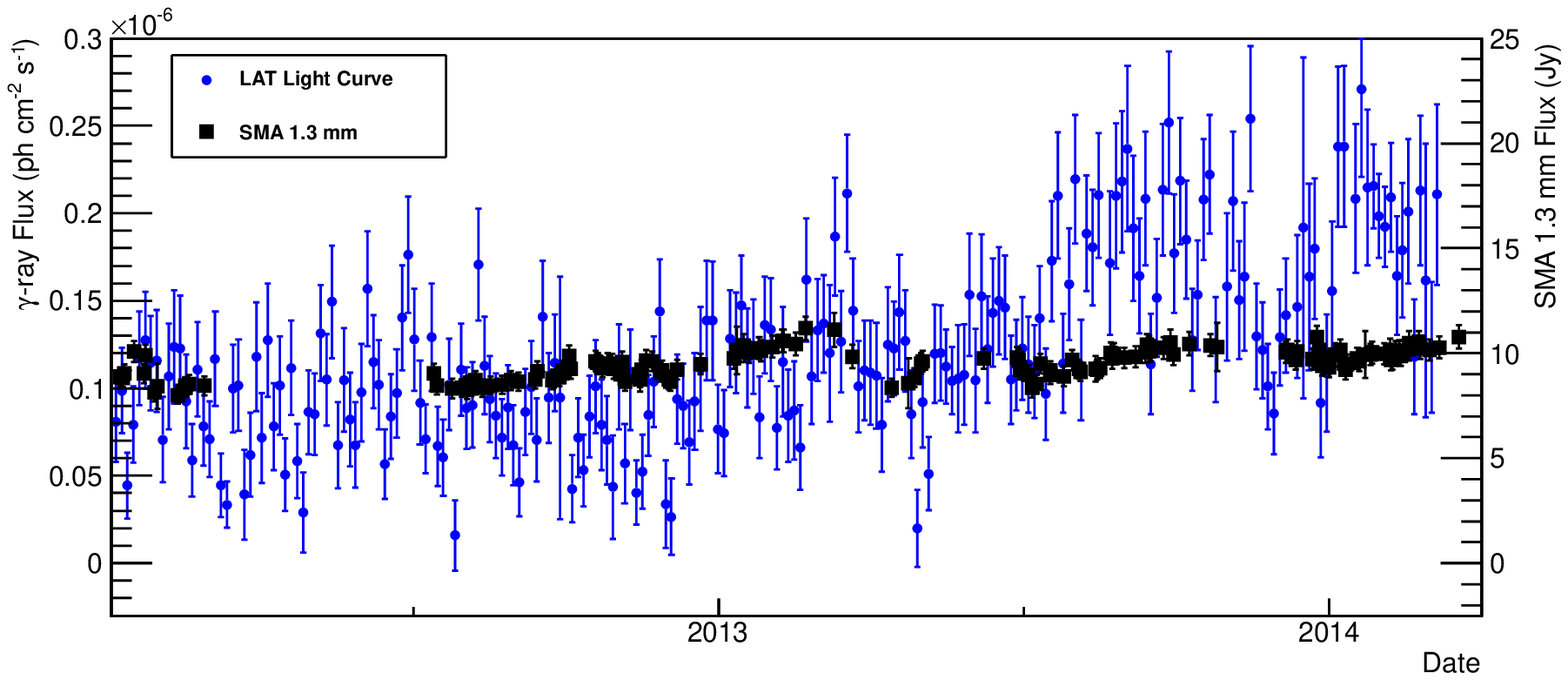}}
  \end{center}
  \caption{Detailed \fermi light curves for NGC\,1275, between
    300\,MeV and 300\,GeV and within 1\degr, focusing on ({\bf top})
    the 2009--2010 period, ({\bf middle}) the 2010--2012 period, and
    ({\bf bottom}) the 2012--2014 period, with 3.5-day binning,
    overlaid with SMA calibrator monitoring data at 1.3\,mm.}
  \label{fig:2}
\end{figure*}

The point-source binned likelihood analysis of NGC\,1275 between the
4$^{th}$ of August 2008 and the 6$^{th}$ of March 2014 outputs an
integrated flux between 300\,MeV and 300\,GeV of
(7.62$\pm$0.10)$\times 10^{-8}$ ph\,cm$^{-2}$s$^{-1}$, a photon index
of 2.141$\pm$0.003, and a Test Statistic of $\sim$27350 (equivalent to a
detection significance of $\sim$165$\sigma$, see \citealp{mattox96}),
where errors are statistical only. Fig.~\ref{fig:1} shows the
fortnightly-binned \fermi light curve for NGC\,1275 over this time
period, together with the model-predicted background flux within the
ROI, and overlaid with the SMA calibrator monitoring data at
1.3\,mm. In the latter, there are several small brightenings
($\sim$\,25--30 per cent) on 1--2 month timescales, but none
comparable to the amplitude observed in 1965--1980 as yet. From the
\fermi light curve, a number of `flares' may be identified, the
earliest of which are discussed in \citet{brown12} and
\citet{kataoka10}. An overall trend of increasing baseline flux is
seen in both datasets.

Fig.~\ref{fig:2} comprises a series of finer (3.5-day binning) \fermi
light curves for NGC\,1275, detailing the 2009--2010, 2010--2012, and
2012--2014 periods (each of which contain prominent flares visible in
the fortnightly-binned light curve), and again overlaid with the SMA
calibrator monitoring data at 1.3\,mm. Though the statistics are poor,
these light curves illustrate the erratic nature of the HE flux level,
highlighting the differing patterns in each flaring event, and
features such as the drop in emission to near-background level in
early 2011 (also apparent in Fig.~\ref{fig:1}).

\begin{figure*}    
  \begin{center}
    \subfigure[][{\label{fig:3a}}]{%
      \includegraphics[width=8cm]{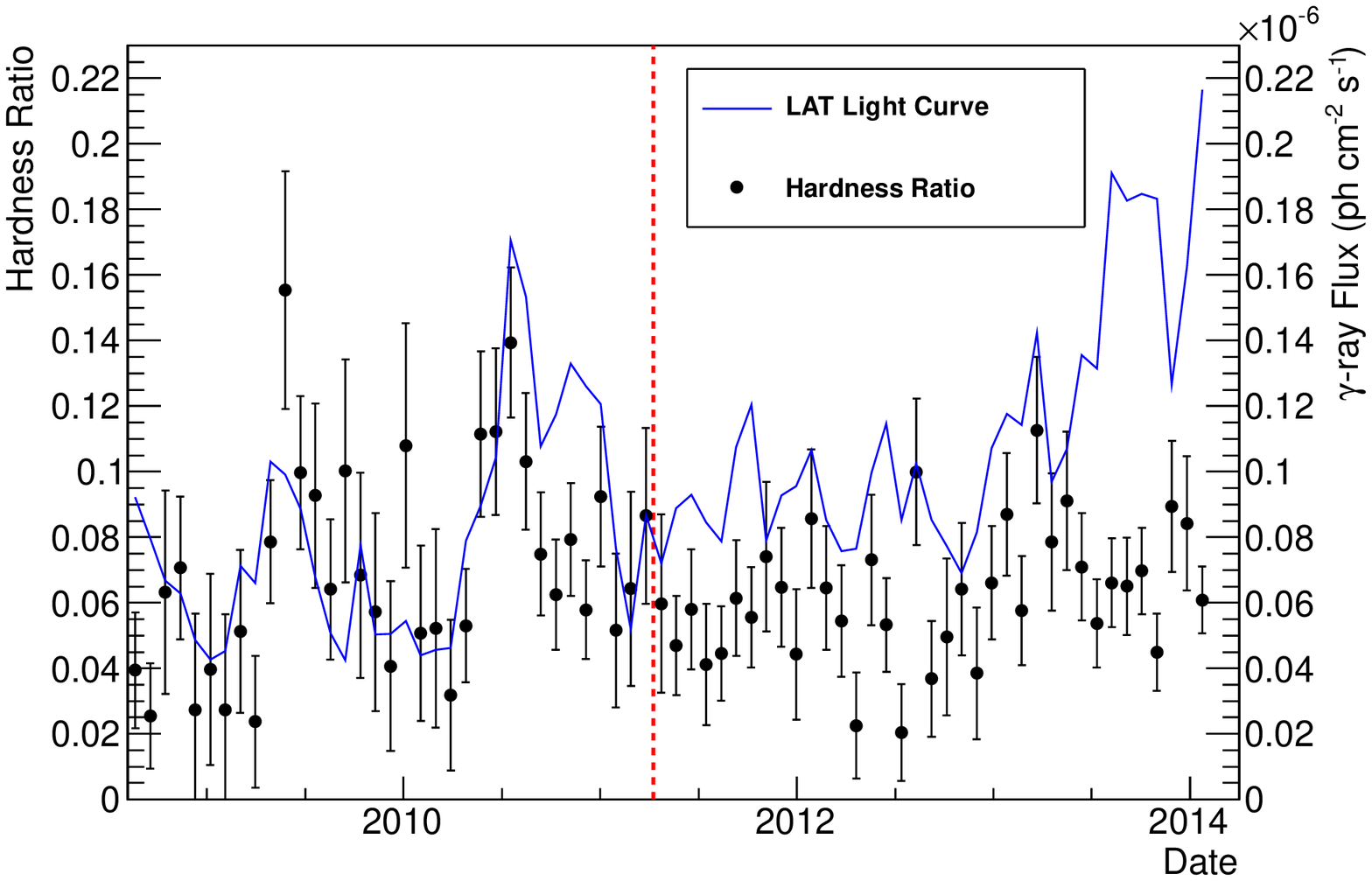}}
    \subfigure[][{\label{fig:3b}}]{%
      \includegraphics[width=8cm]{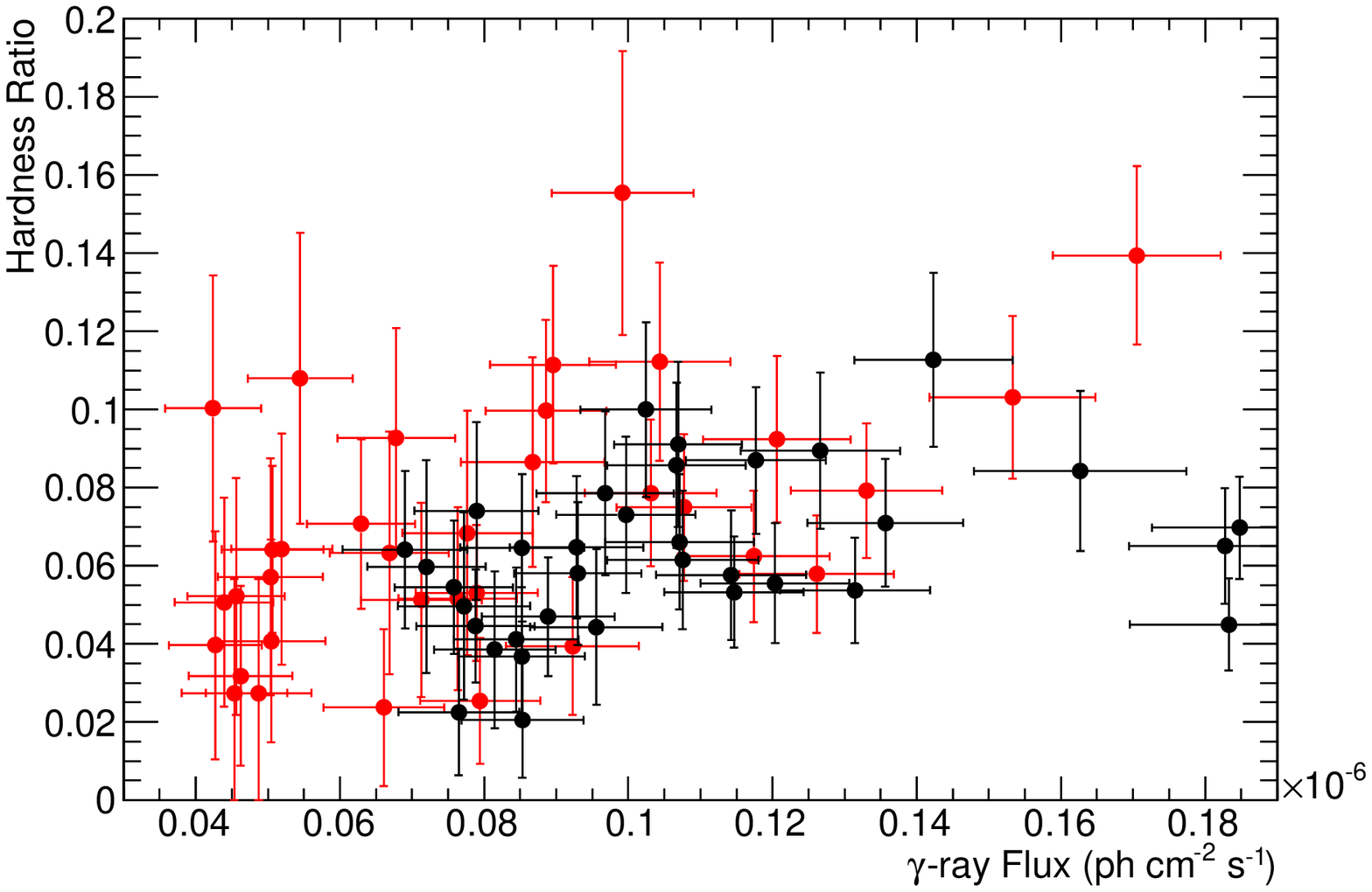}}
  \end{center}
  \caption{A plot of the hardness ratio (defined in
    Section~\ref{sec:analysis}; derived from `high' and `low'-energy
    band data) against time for NGC\,1275, with monthly binning (a),
    and HR plotted against the \gam-ray integrated flux for the full
    energy range for NGC\,1275 (b), with points divided into two time
    periods: prior to (in red) and following (in black) an apparent
    increase in the degree of correlation, as indicated by the dashed
    red line in (a).}
 \label{fig:3}
\end{figure*}

Fig.~\ref{fig:3a} is a plot of the HR against time with monthly
binning, with the source light curve (binned likewise) included for
comparison. The possibility of correlation between HR and flux is
investigated in Fig.~\ref{fig:3b}, which shows the HR plotted against
the integrated flux for the full energy range. Observing harder
emission during brighter periods is in agreement with
\citet{brown12}. There is indeed a hint that during the brightest
months, the corresponding HR is higher, however this correlation is
stronger earlier on, as illustrated by dividing the data into two time
periods: In the thirty-five months following launch, the correlation
factor is 0.53, compared with 0.28 after this epoch, evidencing a
deviation from the established {\it harder-when-brighter} behaviour at
later times.

The light curve-specific ROI does contain another point source of HE
emission (associated with the radio galaxy IC\,310), and this cannot
be easily removed from the light curve, as the emission is barely
resolved from that of NGC\,1275. The background-subtracted light curve
may be very slightly overestimated as a result, but the model, which
considers a larger ROI and SR, does include this source. Likewise
there are several 2FGL sources in the ROI and SR, but these are
included in the model. The two methods of background estimation are in
agreement within the errors associated with the mean background light
curve, which may be fit with a polynomial of order zero, $F_{\gamma} =
(2.50 \pm 0.04)\times 10^{-8}$\,ph\,cm$^{-2}$s$^{-1}$, whilst the
model-predicted background level is $F_{\gamma} = 1.21 \times
10^{-8}$\,ph\,cm$^{-2}$s$^{-1}$. The choice of background estimate
does not very appreciably affect the trend observed in
Fig.~\ref{fig:3b}. A correlation was sought binwise between the
model-subtracted source flux, and the difference between the two
background estimates, but not found, rejecting the hypothesis that
photons from the PSF tail of NGC\,1275 might be contributing
significantly to the flux in the off-regions.  The model-predicted
background was adopted throughout the work presented here, for
consistency within the analysis chain. The aperture correction factor
applied to this background estimate as described in
Section~\ref{sec:analysis} was 1.74 for the overall 300\,MeV$\leq E
\leq$300\,GeV band; 1.32 for the high band and 1.25 for the low band.

\section[]{Discussion}
\label{sec:discussion}

\subsection[]{Historical Consideration}
\label{ssec:history}

\begin{figure*}
  \centering
  \includegraphics[width=16cm]{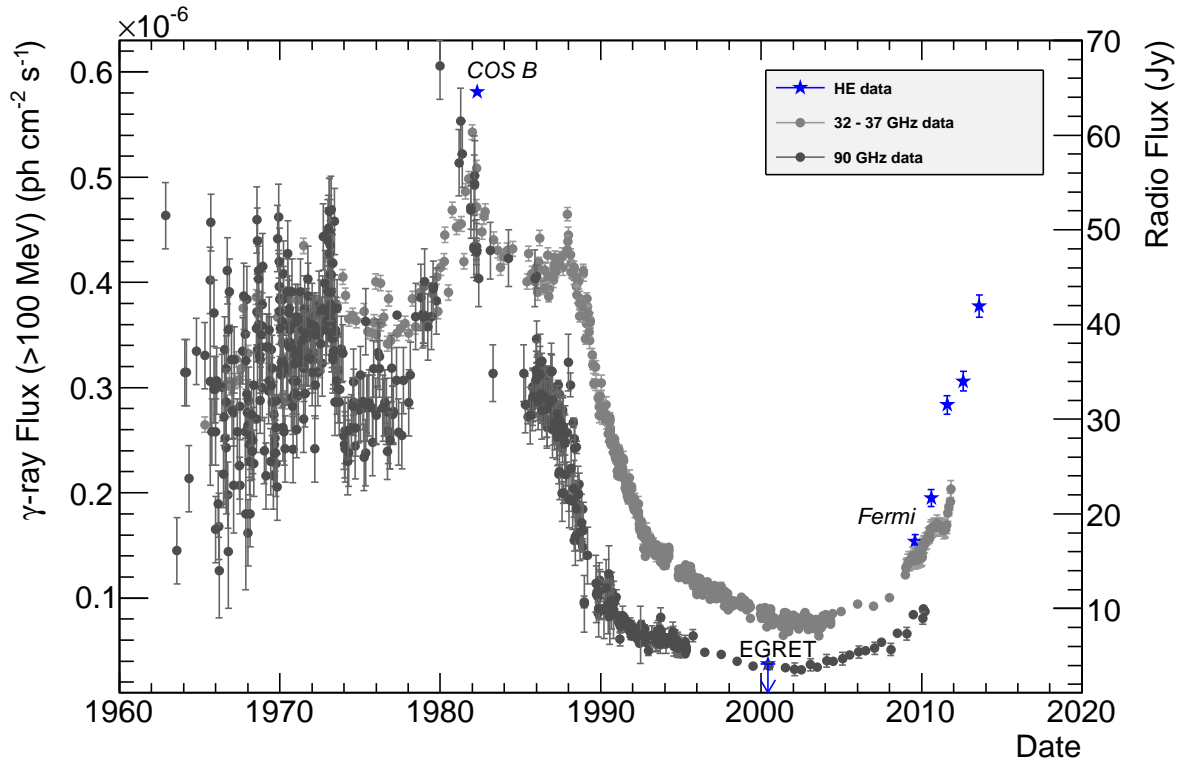}
  \caption{An historical light curve for NGC\,1275. A flux above
    100\,MeV is derived for the emission noted in the {\it COS B} data
    in the direction of the cluster. The EGRET upper limit on the flux
    from Perseus is included, and yearly fluxes for NGC\,1275 as
    observed using the \fermi-LAT. Collected archival radio data
    (described in Section~\ref{sec:analysis} above) is also included,
    at 90 and 32--37\,GHz.}
  \label{fig:4}
\end{figure*}

Historically, NGC\,1275 is a highly-variable source of non-thermal
radiation, strongly motivating its investigation with {\it Fermi}, and
particularly suggestive of a relationship between the HE and radio
emission. Whilst both leptonic and hadronic scenarios are possible for
the \gam-ray emission from NGC 1275, we focus in the following on the
most common interpretation of this emission as inverse Compton
scattering of a relativistic electron population, which is also
responsible for the synchrotron emission seen at radio-mm wavelengths.

The European Space Agency (ESA)'s {\it COS B} mission in the late
1970s \citep{bignami75} revealed an excess of HE emission toward
Perseus ($F_{\gamma}=8.3\times10^{-7}\,{\rm ph}\left (> 70\,{\rm MeV}
\right ){\rm cm}^{-2}{\rm s}^{-1}$; positional uncertainties were not
given), which was interpreted as evidence that NGC\,1275 is a \gam-ray
source \citep[see e.g.][]{strong83}, however the galaxy was not
detected using the National Aeronautics and Space Administration
(NASA)'s Energetic Gamma Ray Experiment Telescope (EGRET) on board the
{\it Compton Gamma-ray Observatory} ({\it CGRO}), which flew
throughout the 1990s. A $2\sigma$ upper limit for the Perseus cluster
of $F_{\gamma} < 3.72 \times 10^{-8}\,{\rm ph}\left (> 100\,{\rm MeV}
\right ){\rm cm}^{-2}{\rm s}^{-1}$ was given
\citep{reimer03}. Contrariwise, NGC\,1275 is firmly-detected using the
      {\it current}-generation \fermi \citep{abdo09a}, and in 2010 was
      detected above 100\,GeV using the Major Atmospheric Gamma
      Imaging Cherenkov (MAGIC) telescopes \citep{aleksic12},
      establishing the source as a VHE emitter. As noted by
      \cite{abdo09a}, the \gam-ray variability of the source matches
      that of the higher-frequency radio emission from the UMRAO
      15\,GHz monitoring. We extend this analysis to 30 and 90\,GHz
      over the past $\sim$50 years, and find that the long-timescale
      variation of both the \gam-ray and millimetre emission are
      related. As shown in Fig.~\ref{fig:1} the connection between
      these two components on shorter timescales is not as
      well-correlated, but this is to be expected if the millimetre
      emission is strongly synchrotron self-absorbed early in any
      flaring event.
      
Combining all of the archival radio data produces 32--37\,GHz and
90\,GHz light curves that show a dramatic variation over the past
$\sim$50 years, as seen in Fig.~\ref{fig:4}. The scatter in photometry
at both frequencies before 1978 may in part be due to the difficulty
in determining accurate fluxes, but could also be related to a period
of enhanced variability, or `flickering', due to larger variations in
accretion during outburst. Interestingly, \citet{nesterov95} note this
period of strong variability in the radio and optical, and speculate
that it is related to the structure of the jet. Taking the period in
which there is significant optical variability as 1965--1980 (from
fig.~1 in \citealt{nesterov95}), the dramatic declines at 90\,GHz and
then at 35\,GHz follow 1--3 and 7--8 years after this,
respectively. Given the increasing time lag seen in the data at lower
frequencies due to synchrotron self-absorption, it may be that the
very gradual decline at $<$100\,GHz is a combination of the many
individual components formed before the 1980 brightening at lower
frequencies, before fading as they move away from the core and become
less self-absorbed.

\begin{figure}
  \centering
  \includegraphics[width=8cm]{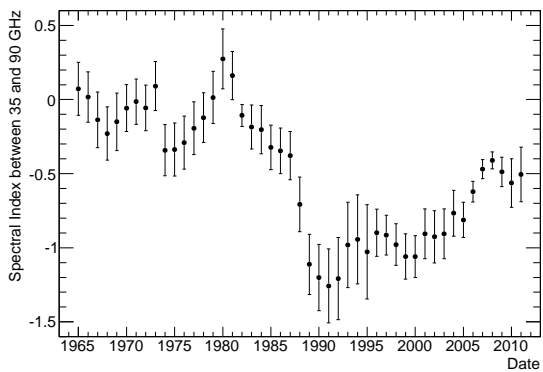}
  \caption{A plot of the average spectral index of NGC\,1275, $\alpha$
    between 35 and 90\,GHz, against time, where the flux is
    proportional to $\nu^{\alpha}$}
  \label{fig:5}
\end{figure}

We have determined the average spectral index between 35 and 90\,GHz
for the period during which the two high frequency light curves
overlap, and show in Fig.~\ref{fig:5} how this spectral index changes
over time. The dramatic transition between the period prior to 1980,
and the 1990--2000 interval, where the spectral index falls from 0.0
to -1.0, is strongly suggestive of a cessation of activity in the core
around 1980, in line with the aforementioned lack of significant
optical variability after that point. The timescale for this change in
spectral index is related to the strength of the synchrotron
self-absorption at different frequencies.

\begin{figure*}    
  \begin{center}
    \subfigure[][{\label{fig:6a}}]{%
      \includegraphics[width=8cm]{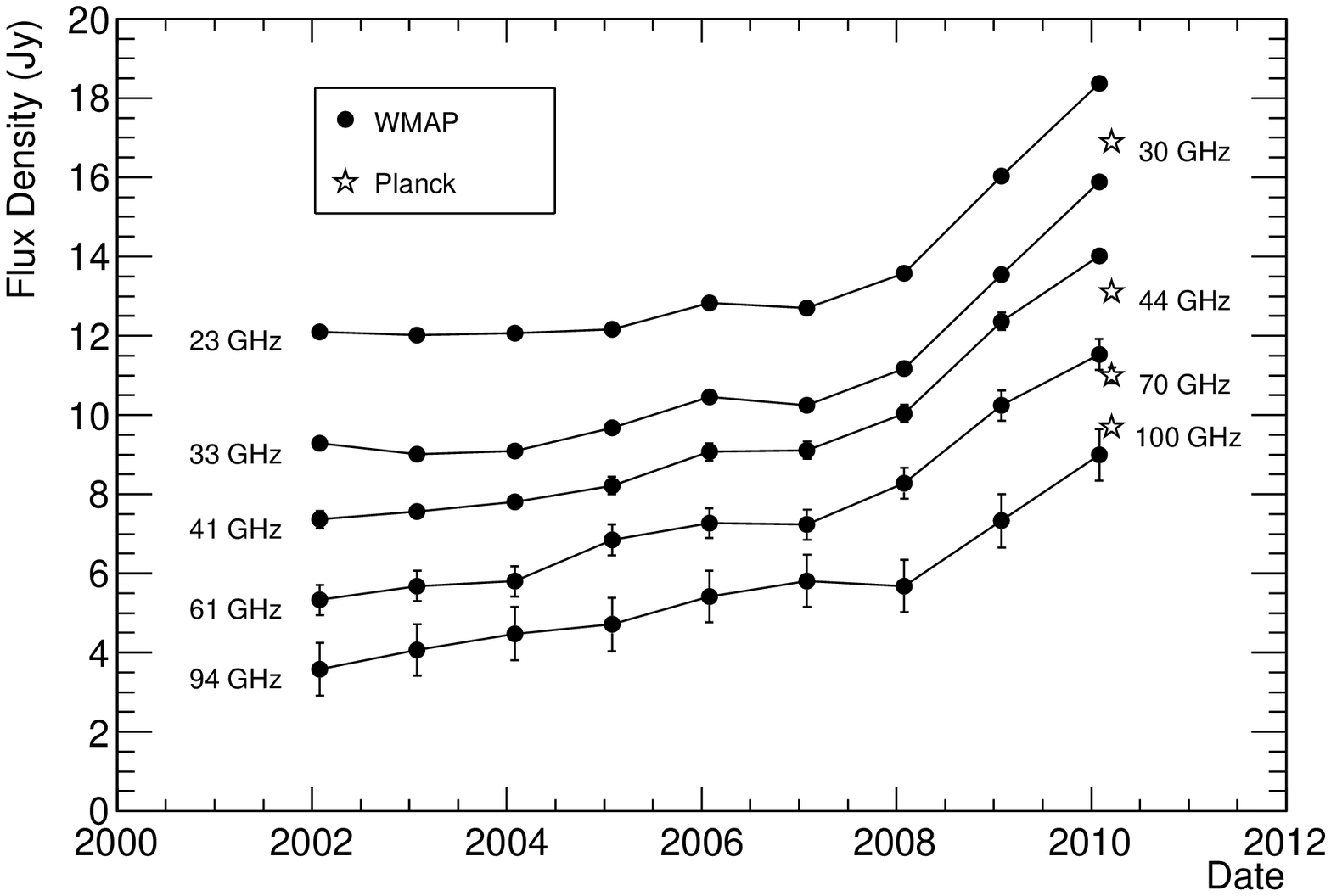}}
    \subfigure[][{\label{fig:6b}}]{%
      \includegraphics[width=8cm]{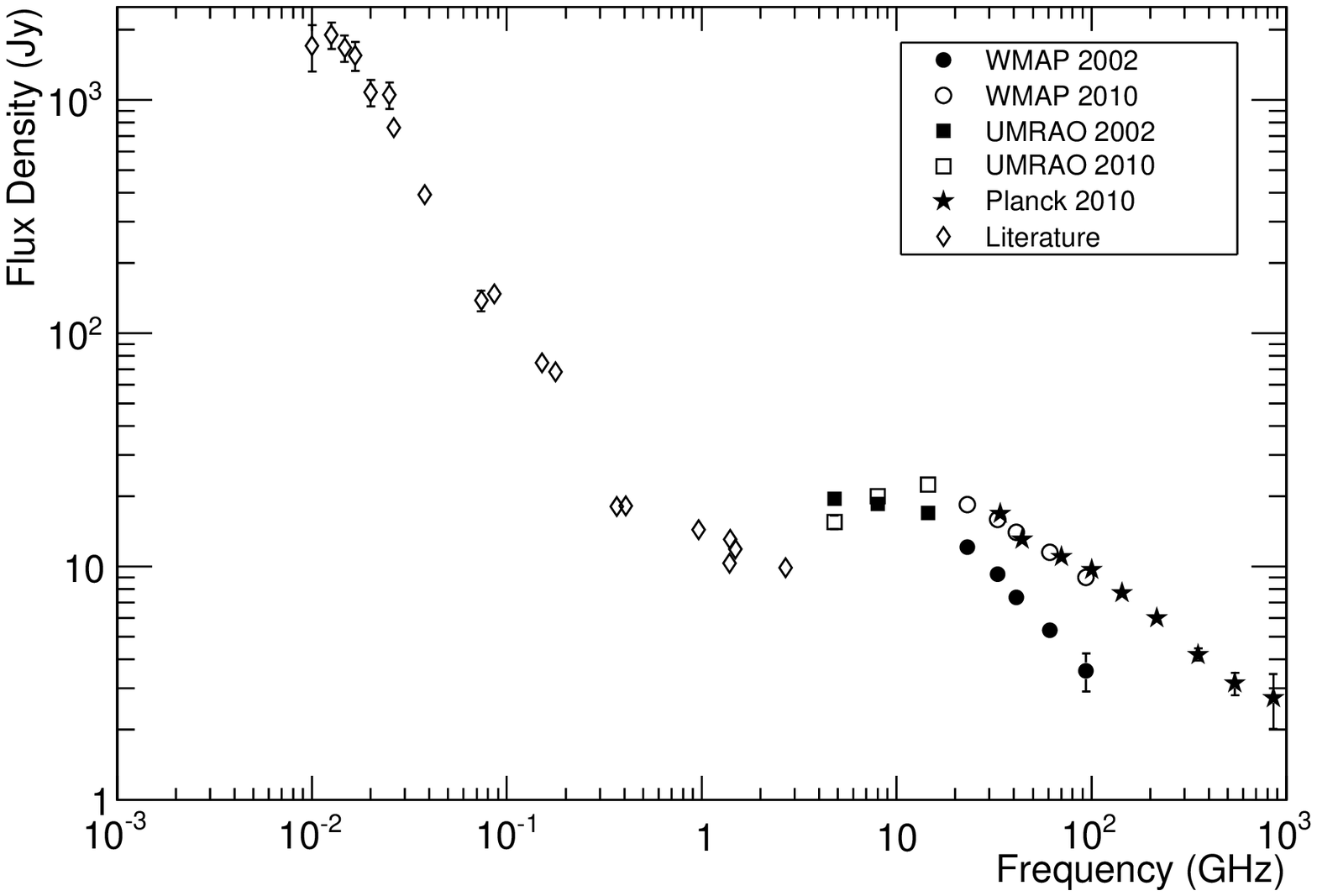}}
  \end{center}
  \caption{Plots showing a recent yearly-averaged light curve (a) and
    the radio-to-far-infrared spectral energy distribution (b) for
    NGC\,1275, using data from {\it WMAP}, {\it Planck}, UMRAO
    monitoring, and the literature.}
 \label{fig:6}
\end{figure*}

Fig.~\ref{fig:6} shows the yearly-averaged {\it WMAP} and {\it Planck}
light curves for the source, and the radio-to-far-infrared spectral
energy distribution (SED) for 2002 and 2010, including points from the
UMRAO monitoring \citep{aller85}, and the literature, to show the
spectral variation at high frequencies ($>$5\,GHz) during this period,
as the flux begins to rise. In particular, note that the 4.8\,GHz
UMRAO flux falls between 2002 and 2010 compared to the increase seen
at all other frequencies.

\subsection[]{Current Outburst}
\label{ssec:current-outburst}

The increase in high-frequency radio and \gam-ray emission from
NGC\,1275, coupled with the change in the spectral index after
$\sim$2002 implies that the source has become active again. The two
epochs in Fig.~\ref{fig:6b} coincide with a minimum in the radio flux,
and a point well into the current outburst, respectively. Importantly,
they mark the start of the {\it WMAP} mission in 2002, and the start
of the {\it Planck} survey in 2009. The well-calibrated photometry,
particularly for {\it Planck}, allows us to determine the spectral
index well above the peak induced by the synchrotron
self-absorption. The most recent data show that the spectral
flattening seen in Fig.~\ref{fig:6b} is related to power law emission
that extends into the far-infrared, and that the core has a more
pronounced GHz-peaked spectrum, with a peak frequency of $\sim$15\,GHz
in 2010.

The complex changes evident from the multi-frequency properties of
NGC\,1275 during the last outburst emphasise how vital regular
monitoring over a wide frequency range (particularly in the mm/sub-mm)
is, if we are to understand the processes that generate this
emission. Fortunately over the past decade, the source has been
regularly observed as part of the F-GAMMA program \citep{angelakis12}
and the Owens Valley Radio Observatory (OVRO) 40\,m Telescope
Monitoring Program \citep{richards11}, and is used as a bright phase
calibrator or pointing source at mm/sub-mm wavelengths by the Institut
de Radioastronomie Millimetrique (IRAM)'s Plateau de Bure
Interferometer (PdBI) \citep{trippe10}, the James Clerk Maxwell
Telescope (JCMT), the SMA, and the Combined Array for Research in
Millimeter-wave Astronomy (CARMA). Therefore, it is very likely that a
well-sampled light curve will be available over the 15--350\,GHz range
to search for any flickering and correlation with the optical and/or
\gam-ray emission.

\begin{figure}
  \centering
  \includegraphics[width=7cm]{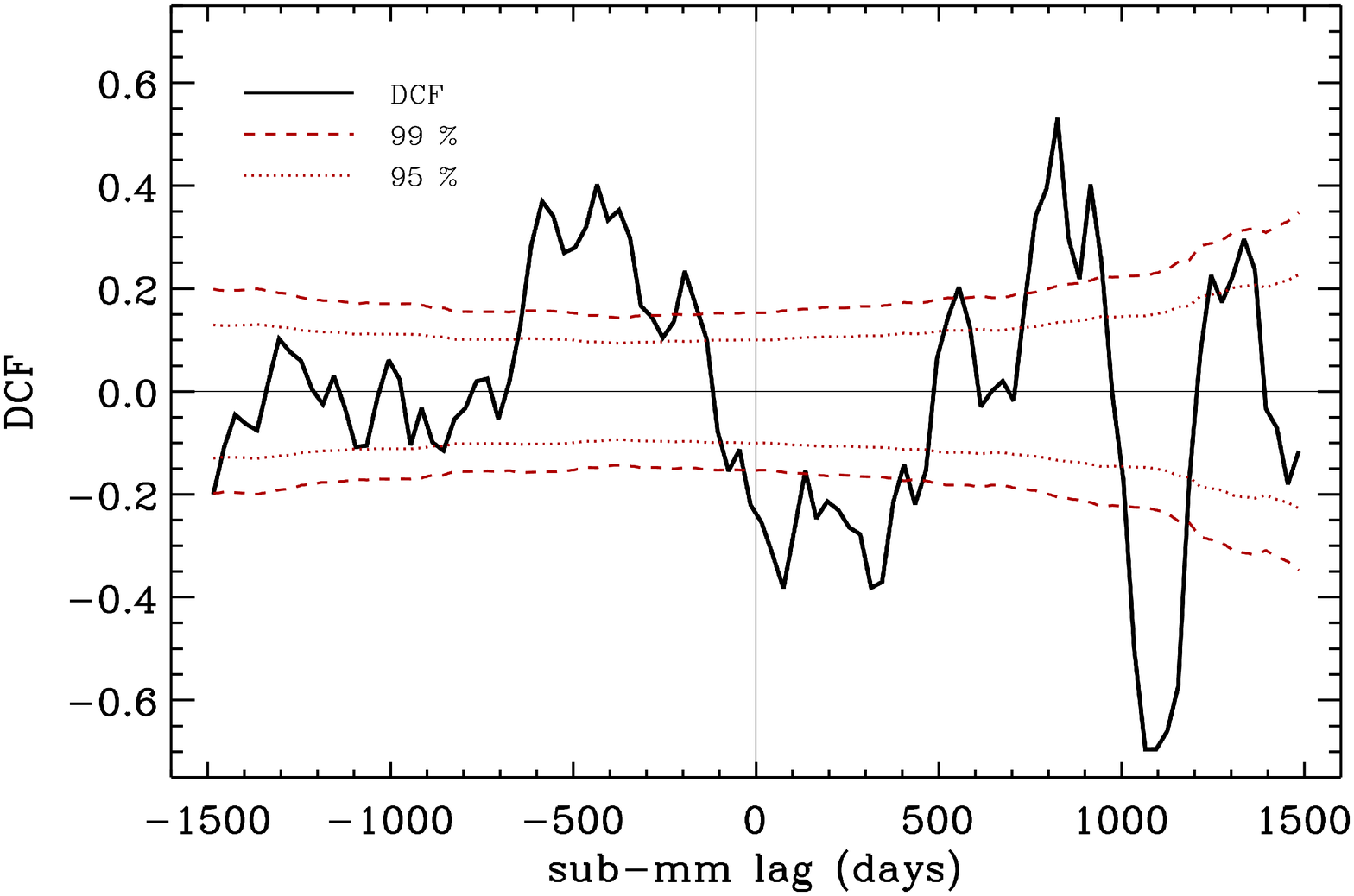}
  \caption{A plot of the DCF produced by cross-correlating the \fermi
    and SMA light curves shown in Fig.~\ref{fig:1}. The 95 and 99 per
    cent confidence intervals estimated using the Bartlett formula
    \citep{bartlett48} are included, though we note that they do not
    include the effect of uncorrelated coloured noise
    \citep[e.g.][]{moerbeck13}.}
  \label{fig:7}
\end{figure}

To illustrate the potential of combining mm/sub-mm monitoring with the
\gam-ray emission we compare the SMA 1.3\,mm data to the fortnightly-
and 3.5-day-binned \fermi light curves (Fig.~\ref{fig:1}
and~\ref{fig:2}). As stated above, the variations at 1.3\,mm do not
coincide with the flares in the \fermi data. Notwithstanding, we
perform a cross-correlation to search for any linear relation between
the two datasets displayed in Fig.~\ref{fig:1}. As the SMA data are
unevenly sampled, we use the Discrete Correlation Function
\citep[DCF;][]{edelson88}. The DCF is only applicable to stationary
data \citep[e.g.][]{welsh99} so a first-order polynomial was first
fitted to, and subtracted from each dataset, in order to remove the
long-timescale, rising trend. The DCF for a range of
-1500$<$lag$<$1500 days is shown in Fig.~\ref{fig:7}. A 30-day DCF bin
width was used (it being larger than the mean sampling rate of the
\gam-ray data of $\sim$15 days), giving a minimum of 110 time lag
estimates in each bin. To assess the significance of any peaks in the
DCF we use the Bartlett formula \citep{bartlett48} to estimate the 95
and 99 per cent confidence intervals, as included in the figure. The
most significant features in the data are a positive correlation at
$\sim +$800 days, as well as a negative correlation seen at $\sim
+$1100 days. A positive correlation is also seen at $\sim -450$
days. Using a DCF bin size of 100 days, the peaks in the DCF are still
present above the 99 per cent level. However, we note that the peaks
in the DCF can arise from the underlying shape of uncorrelated
coloured-noise light curves \citep[e.g.][]{alston13}. Also, removing
both the broad, 5-month peak at the end of 2011 from the HE dataset,
and the 2-month peak in early 2013 from the SMA dataset causes the
sub-mm lag to disappear, which might suggest it is spurious.

The finely-binned light curve illustrated in Fig.~\ref{fig:2} is
suggestive of significant variability on timescales shorter than the
fortnightly binning shown in Fig.~\ref{fig:1}. The possibility of a
flickering nature of the \gam-ray emission is an aspect of the source
that could be used to determine its structure, and so it was
investigated further. We performed chi-square tests with the
variously-binned datasets to determine the scale down to which there
is evidence of variability. Though we found no evidence for
statistically significant variability on the 3.5 day timescale in the
{\it overall} dataset, we found hints of such during shorter,
individual periods. For example, we calculated $\chi^{2}=1.25$ per
degree of freedom for the interval illustrated in the middle panel of
Fig.~\ref{fig:2}, implying a chance probability of consistency with
the trend observed in the fortnightly-binned data of $\sim$1 per cent
It is this period (2010--2012) which contains one of the most
prominent flares. It is clear that on the timescales probed in
Fig.~\ref{fig:2}, evidence for variations from one half-a-week to
another (that do not simply reflect the variability observed in
Fig.~\ref{fig:1}) begins to emerge in some periods, in spite of poor
statistics.

The source of variable high energy \gam~rays is coincident (within
errors), with the nucleus of NGC\,1275. Light-crossing-time arguments
(which state that emitting regions can only fluctuate over a time less
than that taken for light to travel across their extent) mean that any
flux variability observed limits the physical size of (at least some
component of) the source emission. The detection of a \gam-ray flare
by \fermi on timescales of days to weeks places a firm upper limit on
the size of the \gam-ray emitting region of less than about
1000\,AU. On the other hand, the SMA 1.3\,mm monitoring does not show
any strong variation on timescales of less than a week, so this
emission is either self-absorbed, or from a more extended region. The
EGRET upper limit of $F_{\gamma} < 3.72 \times 10^{-8}\,{\rm ph}\left
(> 100\,{\rm MeV} \right ){\rm cm}^{-2}{\rm s}^{-1}$ provides a limit
on the cluster-scale \gam-ray emission of Perseus, which is expected
for example from diffuse populations of long-lived CR hadrons
\citep[see e.g.][]{blasi07}, and in models of dark matter annihilation
\citep[e.g.][]{bringmann12}. Even the lowest flux states evident in
Fig.~\ref{fig:1} are an order of magnitude above this level.

As mentioned in the introduction, 43\,GHz VLBA monitoring of NGC\,1275
shows that one component, `C3', accounts for most of the increase in
the total flux at this frequency
\citep{nagai10,suzuki12,nagai12,aleksic13}. If one assumes that C3 is
coincident with the nucleus then the evolution of the separation
between C1 and C3 presented in \citet{nagai10} and \citet{suzuki12}
can be interpreted as C1 being launched from the nucleus between 2001
and 2003. This coincides with the upturn in activity in the radio and
is a more natural explanation than a rapid change in the apparent
speed of C3 \citep{nagai10}, but implies that the flat spectrum of C1
between 22 and 43\,GHz is not due to it being the core as assumed by
\citet{suzuki12}. This indicates that new VLBA components from flaring
activity in 2010 and 2011 should become visible in the next couple of
years close to C3, and move north if they share the same apparent
velocity as C1 relative to the core. Continued VLBA monitoring of
NGC\,1275 will clarify this issue as it will allow us to link the
complex photometric variations to individual resolved knots.

The lack of any significant flux variation in the SMA 1.3\,mm light
curve during the flares visible in May 2009, August 2010 and September
2011, coupled with the hardening in the \fermi flux at these times,
strongly suggests that whilst the energy of the electrons in the jet
increases and inverse Compton scattering is enhanced during the
flaring periods, the overall electron number density (and hence the
synchrotron emission) is relatively unchanged. Changes in the electron
population; the photons they produce and then up-scatter, depend on
the structure of the emission region. Therefore, variations at one
frequency do not necessarily lead to similar variations elsewhere in
the spectrum.

\citet{zacharias13} demonstrate the time-dependent changes seen in the
overall SED of a source during and after a flare. There is a
significant lag in the X-ray and sub-mm regimes to a flare in the
\gam~rays and optical of days to weeks.  This matches the observation
that NGC\,1275 exhibits enhanced TeV emission in periods of flaring
\citep{aleksic12,aleksic13}. The \citet{zacharias13} models suggest
emission in the optical and \gam~rays will show a more direct link in
their variability. The simultaneous study of \gam-ray and optical
variability has only been possible since the launch of \fermi and
\citet{aleksic13} present the first attempt to investigate this: the
flux of the two datasets are well correlated. They use optical data
from the KVA\footnote{http://users.utu.fi/kani/1m/} optical
monitoring, covering the past four years, which confirms that the
short-timescale ($<$month) variation in optical flux visible in the
\citet{lyutyi77} and \citet{nesterov95} results has returned.

Given the rich variety in the variability and relative brightness of
NGC\,1275, it is to be hoped that it will continue to be monitored at
all wavelengths (especially in the optical and TeV), and more
frequently over the next few years of increasing activity, to search
for direct temporal correlation between the mm/sub-mm, near-infrared,
optical and \gam-ray emission.

\subsection[]{Wider implications}
\label{ssec:implications}

The variable nature of NGC\,1275 has implications for our
understanding of other BCGs in cooling-core clusters. It is known that
the X-ray emission from BCGs is variable \citep[e.g.][]{russell13} and
that a substantial number of GHz-peaked sources are found in BCGs,
both with and without additional, steeper-spectrum components (Hogan
et al., in preparation), but as yet there have been no systematic
searches for radio variability in BCGs in the literature. The example
of NGC\,1275 is an important one, as it highlights the fact that
relying on data from below 10\,GHz can overlook the contribution from
a synchrotron self-absorbed core.

This is particularly important when considering the contamination of
Sunyaev-Zel'dovich (SZ) effect observations in the 30--150\,GHz range,
where the measured decrement could be underestimated if the additional
high-frequency, point source component is not accounted for \citep[see
  e.g.][]{knox04,coble07,lin09}. For instance, the 1.4\,GHz flux
density for NGC\,1275 is comparable to the 100\,GHz {\it Planck} flux
density at $\sim$10--12\,Jy in 2010 but an extrapolation from the
steeper-spectrum emission at lower frequencies would predict a flux of
$\sim$\,0.1\,Jy.  Using NGC\,1275 as an example of such a
contaminating source, it is clear that having radio data over a broad
frequency range and within a few years of the SZ observations is
essential to correctly account for this variable, flat-spectrum core
emission.

We can also use the observed ratio of radio to \gam-ray emission in
NGC\,1275 to estimate the possible \fermi fluxes of other
BCGs. Scaling from the 2010 fluxes at 90\,GHz and between
0.3--300\,GeV of 10\,Jy and 3$\times 10^{-8}$ph\,cm$^{-2}$\,s$^{-1}$,
respectively, the next brightest BCG in the radio band would be
50--100 times fainter in the \gam~rays, so undetectable at the current
\fermi exposure level. This agrees with the analysis of
\citet{dutson13} who considered a sample of 114 core-dominated BCGs
and found no individual \fermi detections apart from two AGN-dominated
sources, already established as HE emitters. Nor was a signal found by
co-adding the BCG candidates, so it is uncertain whether it is the
proximity of NGC\,1275 that makes it exceptional, or that it is in
itself peculiar. Its current state is not necessarily typical of the
source; continued monitoring decades into the future might address
this uncertainty, establishing observationally a complete cycle into
and out of an active state. If another moderately bright BCG were to
experience an outburst, then \fermi ought to be capable of detecting
it (particularly during flaring events) but it would require long-term
monitoring of dozens of systems at $>$30\,GHz to identify possible
candidates.

The properties of NGC\,1275 can also inform our understanding of radio
galaxies and BL\,Lac objects. The beaming of the core that makes
BL\,Lacs so prominent also acts to shorten the observed variability
timescales at all wavelengths, by an order of magnitude in many
cases. Therefore, the outburst timescale observed in NGC\,1275 of a
few decades with at most {\it mild} beaming \citep{nagai10}, is
comparable to the few-year outburst timescale seen in sources such as
3C\,279 and 3C\,454.3, where the beaming is a factor of order ten
higher. Thus, when considering the core dominance of radio galaxies it
is important to take into account both the self-absorption of the
core, and its potential variability on decade-to-century
timescales. The observed properties of NGC\,1275 show very little
variation below 1\,GHz, so selecting a sample of radio galaxies at low
frequency (e.g. 159\,MHz using the 3C Catalogue; \citealp{edge59})
would select sources like NGC\,1275 uniformly, as the level of core
activity has little or no effect. However, making the selection at
$>15$\,GHz (e.g. The Australia Telescope 20\,GHz Survey; AT20G;
\citealp{murphy10}) would depend greatly on when in the last 50 years
the object was observed. Ensuring that the `time domain' aspects of
the properties of radio galaxies are considered will help build a more
self-consistent picture of both beamed and non-beamed radio sources,
and the emission mechanisms that give rise to their extraordinary
multiwavelength properties.

\section[]{Concluding Remarks}
\label{sec:conclusion}

NGC\,1275 has been pivotal to our understanding of AGN feedback over
the past three decades \citep[see e.g.][]{bohringer93,fabian00}. The
presence of multiple `bubbles' in the intracluster medium of the
surrounding cluster indicates that there are major outbursts from the
core on timescales of 10$^{6-7}$\,years.

Building on previous work, this paper provides striking evidence for
variability of NGC\,1275 on decade timescales, showing that there are
repeated outbursts from the central few parsecs of the source, which
vary in amplitude by about an order of magnitude.  On top of this
episodic variability there is also short-timescale flaring activity
seen over a few days to a month in the \gam~rays, that can produce an
additional order-of-magnitude brightening in flux and coincide with a
detection at TeV energies.

NGC\,1275 is currently brightening rapidly and may reach a luminosity
similar to that observed for over a decade in the 1960's and
1970's. The much broader range of multi-wavelength facilities
available now will present an unprecedented opportunity to study the
behaviour of the source during a very active phase. In particular,
linking the shorter-timescale \gam-ray variability to the launching of
individual radio components in the jet \citep[e.g.][]{suzuki12} is now
possible with continued monitoring with long baseline interferometers
at high frequency ($>$20\,GHz). Therefore, the current outburst will
provide us with an opportunity to witness processes at the very heart
of AGN feedback.

\section*{Acknowledgements}

The SMA is a joint project between the Smithsonian Astrophysical
Observatory and the Academia Sinica Institute of Astronomy and
Astrophysics, and is funded by the Smithsonian Institution and the
Academia Sinica. This work has made use of public \fermi data and
Science Tools provided by the \fermi Science Support Centre, the
NASA/IPAC Extragalactic Database (NED) which is operated by the Jet
Propulsion Laboratory, California Institute of Technology, under
contract with NASA, and the SIMBAD and VIZIER databases, operated at
CDS, Strasbourg, France. We thank Margo Aller for permission to use
data from the University of Michigan Radio Astronomy Observatory,
which was supported by the National Science Foundation and NASA, and
by funds from the University of Michigan. We thank also Richard White,
whose automated \fermi tool chain improved the time-efficiency of much
of the analysis carried out. K.~L. Dutson acknowledges support from
the STFC studentship ST/I505780/1 and the STFC Doctrinal Training
Grant ST/J501104/1, M.~T. Hogan from the STFC studentship
ST/I505656/1, W.~N. Alston from the EU FP7-SPACE research project,
STRONGGRAVITY., J.~A. Hinton from the Leverhulme Trust, and A.~C. Edge
from STFC grant ST/I001573/1.

\bibliography{Master}
\bibliographystyle{mn2e}

\appendix
\bsp
\label{lastpage}

\end{document}